\documentstyle[sprocl,epsfig]{article}

\def\be{\begin{equation}}
\def\ee{\end{equation}}
\def\bea{\begin{eqnarray}}
\def\eea{\end{eqnarray}}

\newcommand{\ga}{\alpha}

\newcommand{\gs}{\sigma}

\newcommand{\gD}{\Delta}

\begin{document}
\title{Quartic gauge couplings from $e^+e^-\rightarrow W^+W^-Z$}
\author{M. Beyer, S. Christ, E. Schmidt, H. Schr\"oder} \address{FB
  Physik, University of Rostock, 18051 Rostock, Germany}
\maketitle

Effective quartic gauge couplings on tree level appear in a strong
interacting Higgs scenario, which could be relevant, e.g. if no
standard model Higgs would be found~\cite{Kilian:2003pc}. This
scenario can be described in terms of an effective chiral Lagrangian
valid for invariant energies $m_{W,Z}<\sqrt{s}<4\pi v$, where the
vacuum expectation value due to electroweak symmetry breaking is
$v=(\sqrt{2} G_F)^{-1}\simeq 246$ GeV. It can be tested in a future
linear collider~\cite{TDR}. The Higgs potential is
given as in the standard model
and the Higgs fields can be described by a nonlinear realization of
chiral symmetry~\cite{Donoghue:1992dd}.
The spontaneous symmetry breaking leads to masses of the gauge bosons.
A systematic one-loop low energy expansion in the energy regime
considered here leads to counter terms that require parameters that
have to be determined by experiments. Because of the equivalence
theorem the Higgs fields can be identified with the longitudinal
components of the gauge bosons. Hence these counter terms can be
identified with effective (anomalous) couplings between the gauge
bosons. The generic quartic terms are~\cite{Kilian:2003pc}
\begin{equation}
{\cal L}_4 = \alpha_4 ({\rm tr}V_\mu V_\nu)^2 \qquad\qquad
{\cal L}_5 = \alpha_5 ({\rm tr}V_\mu V^\mu)^2
\end{equation}
with $
V_\mu\equiv-i\frac{g_W}{\sqrt{2}}(W_\mu^+\tau^++W_\mu^-\tau^-)
-ig_ZZ_\mu\frac{\tau_3}{2}$.
This interactions can be investigated trough weak $WW$
fusion~\cite{Hagiwara:1991zz}. The sensitivity at TESLA energies has
been studied in Refs.~\cite{TDR,Boos:1999kj}. A comparison between
$WW$ fusion and $WWZ$ or $ZZZ$ production at LHC energies is given in
Ref~\cite{Han:1997ht}. Recently $WW$ fusion has been investigated for
the $e^-e^-$ option~\cite{monig04}. Induced quartic coupling via
triple gauge coupling will not been considered here. Triple gauge
couplings have been investigated previously~\cite{TDR}.

\begin{figure}[t]
\begin{minipage}{0.47\textwidth}
\begin{center}
\epsfig{figure=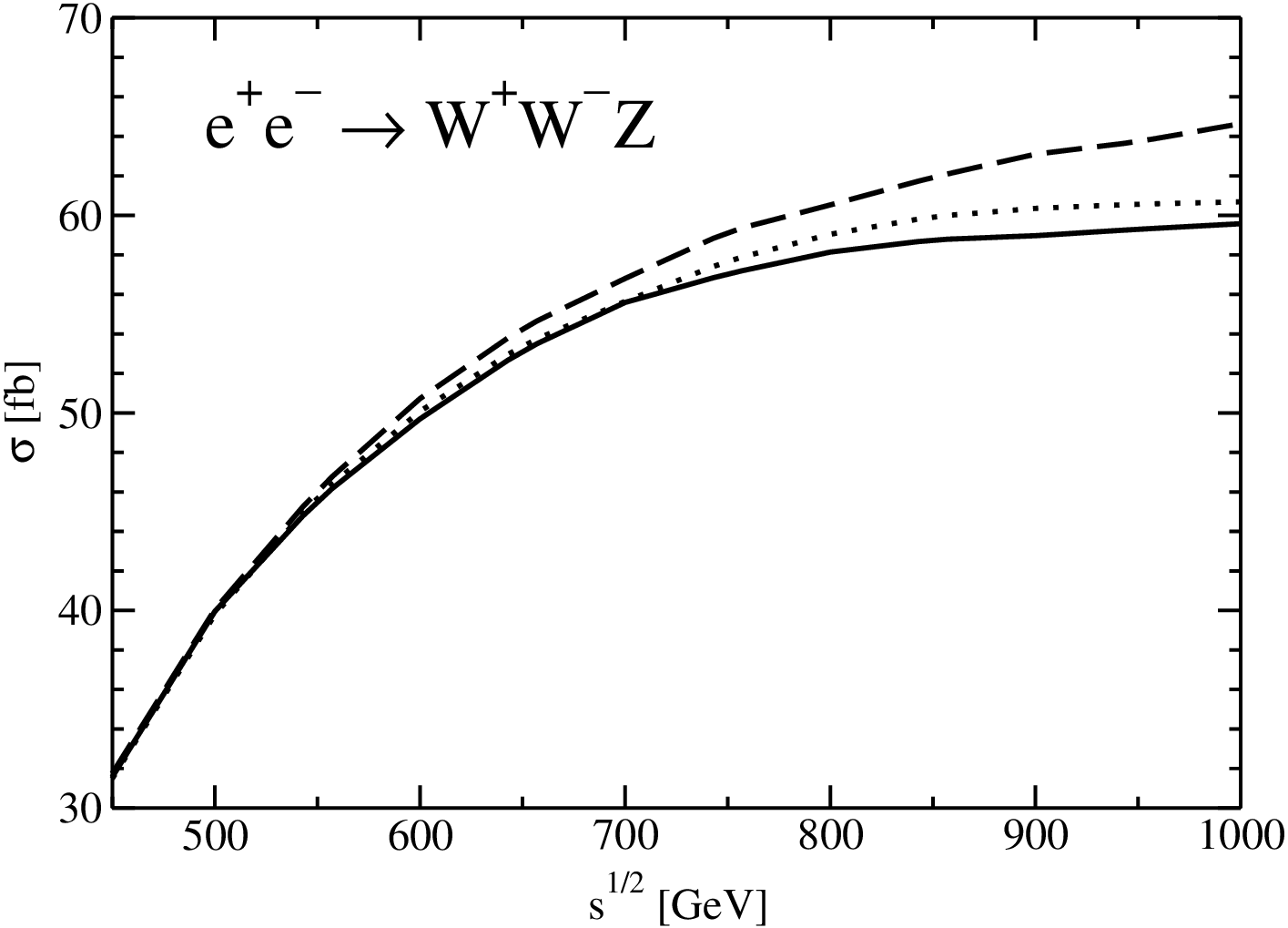,width=\textwidth}
\end{center}
\caption{\label{fig:cross_s} Energy dependence of partial cross section:
  standard model (solid line), $\ga_4=0.1$, $\ga_5=0$ (dotted),
  $\ga_4=0$, $\ga_5=0.1$ (dashed).\hfill~ }
\end{minipage}\hfill
\begin{minipage}{0.47\textwidth}
\begin{center}
\epsfig{figure=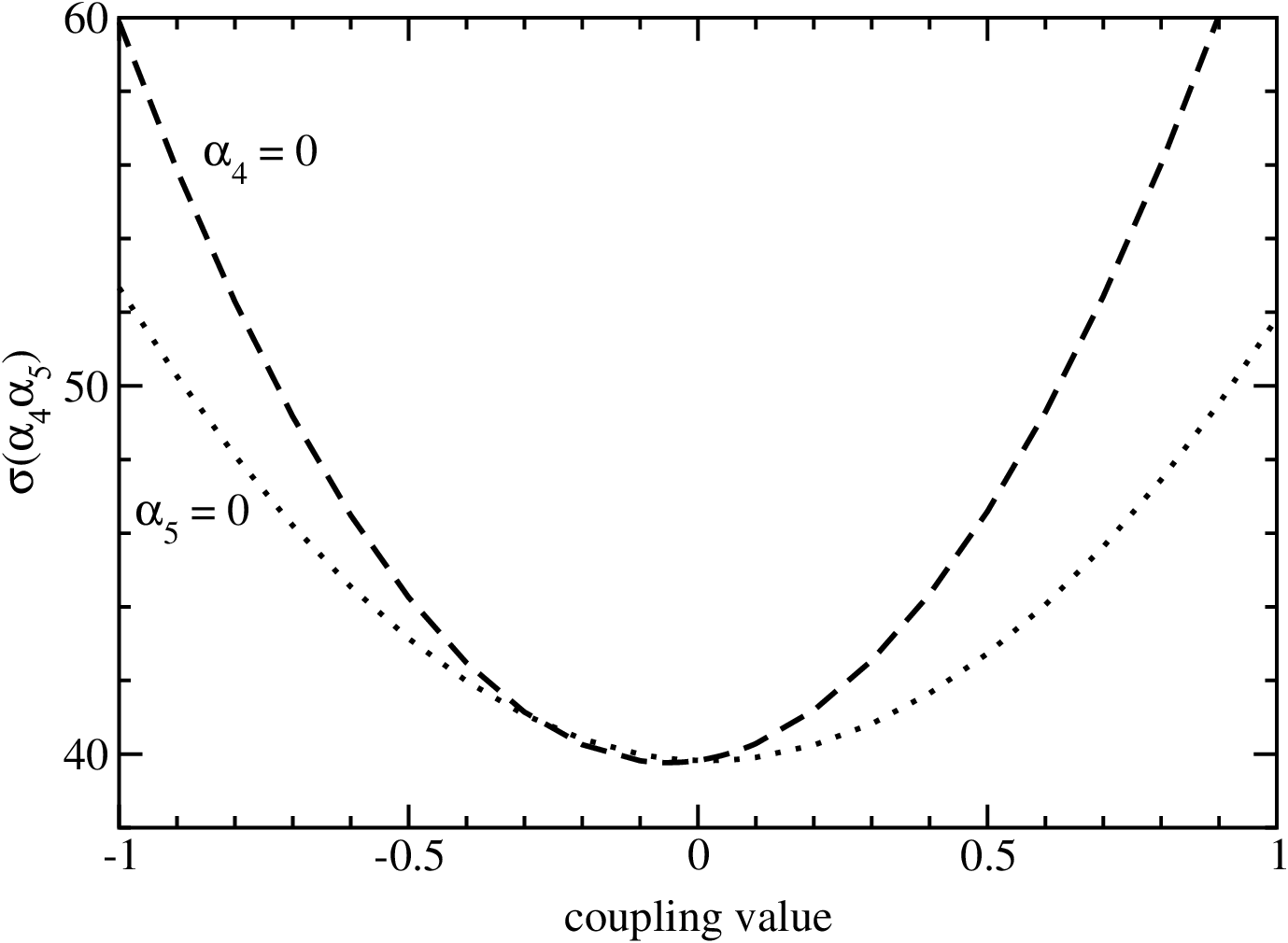,width=\textwidth}
\end{center}
\caption{\label{fig:cross} Dependence of cross section 
  on $\ga_4$ and $\ga_5$ at $\sqrt{s}=500$ GeV. Standard model value
  for $\ga_4=\ga_5=0$ other $\ga_4=0.1$, $\ga_5=0$ (dotted),
  $\ga_4=0$, $\ga_5=0.1$ (dashed).\hfill~}
\end{minipage}
\end{figure}

As an example we presently consider the reaction $e^+e^-\rightarrow
W^+W^-Z$. On tree level the elementary process is driven by 15 Feynman
diagrams.  Only one of the diagrams contains the quartic coupling and
has to be extracted from the other interfering terms. Furthermore only
the part containing longitudinal gauge bosons is expected to give a
sizable signal related to electroweak symmetry breaking. Since the
gauge bosons are short living states they decay and off-shell effects
have to be taken into account. These can be accommodated by
considering 6 fermion final states on the parton level, which is
possible using Whizard as an event generator~\cite{Kilian:2001qz}.
Presently we consider on-shell gauge bosons only (narrow width
approximation) and hadronize the final state using
PYTHIA~\cite{Sjostrand:2001yu}. This restriction, however, will be
relaxed as the investigation goes on.  The partial cross section for
the (no-Higgs) standard model result is shown as a solid line in
Fig.~\ref{fig:cross_s}. The other lines include the contributions from
$\ga_4=0.1$ and $\ga_5=0.1$. The total cross section at 500 GeV as a
function of $\ga_4$ or $\ga_5$ is shown in Fig.~\ref{fig:cross}.

The three-boson state $WWZ$ is characterized by three four-momenta and
the spins. In general three momenta lead to 12 kinematical variables
that are reduced by 4 through energy momentum conservation, by 3
because of the on-shell condition mentioned before, and by 2 due to
rotational invariance. Hence in total three independent kinematical
variables are left. We choose two invariant masses of the Dalitz plot,
$M_{WZ}^2$, $M_{WW}^2$ and the angle $\theta$ between the beam axis
and the direction of the $Z$-boson. Spin leads to additional degrees of
freedom, and we may differ between longitudinal and transverse
polarization of the bosonic spins. Presently, we do not analyze the
spins of the bosons.

\begin{figure}[t]
\begin{minipage}{0.45\textwidth}
  \epsfig{figure=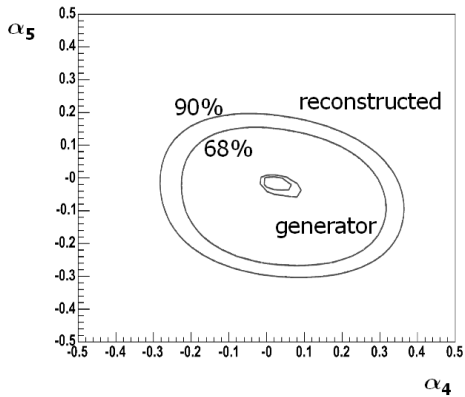,width=\textwidth, angle=270} %
\end{minipage}
\hfill
\begin{minipage}{0.40\textwidth}
\caption{\label{fig:contur}
Expected sensitivity for $ga_4$ and $\ga_5$ at $\sqrt{s}=500$
  GeV for the process $e^+e^-\rightarrow W^+W^-Z$. The two inner lines
  represent the results on generator level (statistical error from 2
  million events, 68\% and 90\% confidence). The outer two lines
  include detector efficiency and purity for a luminosity of 1000
  fb${-1}$. To compare with $2\rightarrow 2$ of the TDR the axis have
  to be scaled by $16\pi^2$.}  
\end{minipage} 
\end{figure}
The three independent kinematical variables lead to a three
dimensional histogram. If the angle $\theta$ is not measured the
resulting two dimensional histogram leads a Dalitz plot. We
investigate the differences on the histograms as a function of $\ga_4$
and $\ga_5$. The observables are discretized into bins and $\chi^2$ is given by
\begin{equation}
\chi^2=\sum_{i,j,k}\frac{N^{\rm exp}_{ijk}-N^{\rm theo}_{ijk}(\ga_4,\ga_5)}
{\gs_{ijk}^2}
\end{equation}
where $\gs_{ijk}$ denotes the error, and $i,j,k$ the sums over bins of
$M_{WZ}^2$, $M_{WW}^2$, and $\theta$. The theoretical values are
achieved using Whizard that provides the partonic cross section and
an interface to PYTHIA to do the hadronization. The detector
efficiency is simulated using the fast simulation
SIMDET~\cite{Pohl:2002vk}. We used a total number of 2
million events. In lack of experimental values we use as a reference an
independent ensemble of 50,000 standard model events, that adds up to
a luminosity of 1,000 fb$^{-1}$ for $\sqrt{s}=500$ GeV. Since the
effective Lagrangian is linear in $\ga_4$, $\ga_5$ any observable is
of second order in the parameters.  Hence, we evaluate $N^{\rm
  theo}_{ijk}(\ga_4,\ga_5)$ for a number of pairs $(\ga_4,\ga_5)$ (15
to be specific) and fit
\begin{equation}
N^{\rm  theo}_{ijk}(\ga_4,\ga_5)=
N_{ijk}^{\rm sm}+N_{ijk}^{\rm A}\ga_4+
N_{ijk}^{\rm B}\ga_4^2+N_{ijk}^{\rm C}\ga_5+
N_{ijk}^{\rm D}\ga_5^2+N_{ijk}^{\rm E}\ga_4\ga_5
\end{equation}
for each bin $i,j,k$ by adjusting the 6 parameters $N_{ijk}^{\rm
  sm},\dots,N_{ijk}^{\rm E}$ using Minuit. Finally we calculate
$\chi^2$ and determine $\gD\ga_4(\ga_4,\ga_5)$ and
$\gD\ga_5(\ga_4,\ga_5)$ for the specific values $\chi^2=2.30$ (68.3\%
confidence) and $\chi^2=4.61$ (90\% confidence). The results are
depicted in Fig.~\ref{fig:contur}.

To reconstruct the $WWZ$ from the detector (SIMDET), we use the decay
of $WWZ\rightarrow 6$ jets. The branching ratio is about 32\% which is
about 16,000 events for the kinematics considered. The dominant
background is due to $t\bar t\rightarrow b\bar b WW\rightarrow 6$ jets
which adds up to 220,000 events. Events selection is done in the
following way. We enforce 6 jets with missing energy and momentum
$E_{\rm mis}^2+p_{\perp,\rm mis}^2 <(65 {\rm GeV})^2$ and a minimum
jet energy of $E_{\rm min}^{\rm jet}>5$ GeV. To reconstruct the $WWZ$
we arrange the 6 jets into all possible 3 pairs, determine the
candidate mass.  We require $|m^{\rm cand}-m^{\rm true}|< 10$ GeV
(where $m^{\rm true}$ is taken from the Particle Data Group), and take
the best combination. To identify the dominant background we combine a
candidate $W$ from $t\rightarrow bW$ with 1 jets assume $|m^{\rm
  cand}_t-m^{\rm true}_t|< 15$ GeV and require an event consistent
with the $t\bar t$ topology. This leads to an efficiency of about 60\%
and a purity of the signal of about 70\%.

\section*{Conclusion and Perspectives} 

In the present analysis the potential of the process considered here
has not yet been fully explored. So far we have considered
correlations between two kinematical variables. Preliminary
exploration shows that the sensitivity can be enhanced by about a
factor of five, if the angle $\theta$ is considered in addition.
Further improvements are possible, if longitudinal gauge bosons only
are used in the analysis. Besides dominant 6-jet events, 4 jets
$\ell^+\ell^-$ play an important role, because they provide rather
clean signals.  Furthermore $ZZZ$ events have not been considered so
far. The standard model cross section at 500 GeV is about 0.87
fb$^{-1}$, i.e. much smaller than the corresponding $WWZ$ channel.
However, the $ZZZ$ channel is also sensitive to other anomalous
couplings than the ones considered here. The narrow width
approximation can be relaxed by using the full potential of Whizard
that enables us to treat 6 fermion final states. From
Fig.~\ref{fig:cross_s} is it clear that sensitivity to the anomalous
couplings should be larger at higher energies, which is the next step
in the analysis.\\
{\em Acknowledgement.} We like to thank Wolfgang Kilian for assistance in
 Whizard et al.

\end{document}